\documentclass[aps,prl,twocolumn,superscriptaddress]{revtex4-2}
\usepackage{amsmath,amssymb,wasysym,graphicx}

\usepackage[utf8]{inputenc}
\usepackage[T1]{fontenc}
\usepackage{xcolor}

\IfFileExists{newtxtext.sty}
{\usepackage{newtxtext,newtxmath}}
{\IfFileExists{stix.sty}
	{\usepackage{stix}}
	{\IfFileExists{mathptmx.sty}
		{\usepackage{mathptmx}}{} } }

\usepackage{textcomp}

\usepackage{bm}

\IfFileExists{siunitx.sty}{\usepackage{booktabs,siunitx}}{}

\usepackage{color}
\definecolor{LinkColor}{rgb}{0.256,0.439,0.588}
\usepackage{hyperref}
\hypersetup{
	pdfauthor={good guys},
	pdftitle={good title},
	colorlinks=true,
	citecolor=LinkColor,
	linkcolor=LinkColor,
	urlcolor=LinkColor
}

\usepackage{pifont}
\usepackage[normalem]{ulem}

\newcommand{\RNum}[1]{\uppercase\expandafter{\romannumeral #1\relax}}

\usepackage{multirow}

\begin{document}

\title{Evidences for the random singlet phase in a new honeycomb iridate SrIr$_2$O$_{6-\delta}$}
\author{Pengbo Song}
\affiliation{Beijing National Laboratory for Condensed Matter Physics, Institute of Physics, Chinese Academy of Sciences, Beijing 100190, China}
\affiliation{Center of Materials Science and Optoelectronics Engineering, University of Chinese Academy of Sciences, Beijing 100049, China}
\affiliation{School of Physical Sciences, University of Chinese Academy of Sciences, Beijing 100190, China}
\author{Kejia Zhu}
\affiliation{School of Physics and Materials Science, Anhui University, Hefei, Anhui, 230601 P. R. China}
\affiliation{Beijing National Laboratory for Condensed Matter Physics, Institute of Physics, Chinese Academy of Sciences, Beijing 100190, China}
\author{Fan Yang}
\affiliation{School of Physics, Beihang University, Beijing 100191, China}
\author{Yuan Wei}
\affiliation{Beijing National Laboratory for Condensed Matter Physics, Institute of Physics, Chinese Academy of Sciences, Beijing 100190, China}
\affiliation{School of Physical Sciences, University of Chinese Academy of Sciences, Beijing 100190, China}
\author{Lu Zhang}
\affiliation{Beijing National Laboratory for Condensed Matter Physics, Institute of Physics, Chinese Academy of Sciences, Beijing 100190, China}
\affiliation{School of Physical Sciences, University of Chinese Academy of Sciences, Beijing 100190, China}
\author{Huaixin Yang}
\affiliation{Beijing National Laboratory for Condensed Matter Physics, Institute of Physics, Chinese Academy of Sciences, Beijing 100190, China}
\affiliation{School of Physical Sciences, University of Chinese Academy of Sciences, Beijing 100190, China}
\affiliation{Yangtze River Delta Physics Research Center Co., Ltd., Liyang, Jiangsu, 213300, People’s Republic of China}
\author{Xian-Lei Sheng}
\affiliation{School of Physics, Beihang University, Beijing 100191, China}
\author{Yang Qi}
\affiliation{Center for Field Theory and Particle Physics, Department of Physics, Fudan University, Shanghai 200433, China}
\affiliation{State Key Laboratory of Surface Physics, Department of Physics, Fudan University, Shanghai 200438, China}
\author{Jiamin Ni}
\affiliation{State Key Laboratory of Surface Physics, Department of Physics, Fudan University, Shanghai 200438, China}
\author{Shiyan Li}
\affiliation{State Key Laboratory of Surface Physics, Department of Physics, Fudan University, Shanghai 200438, China}
\author{Yanchun Li}
\affiliation{Multidiscipline Research Center, Institute of High Energy Physics, Chinese Academy of Science, Beijing 100049, China}
\author{Guanghan Cao}
\affiliation{Department of Physics, Zhejiang University, Hangzhou 310027, China}
\affiliation{Zhejiang Province Key Laboratory of Quantum Technology and Devices, Interdisciplinary Center for Quantum Information, and State Key
Lab of Silicon Materials, Zhejiang University, Hangzhou 310027, China}
\affiliation{Collaborative Innovation Centre of Advanced Microstructures, Nanjing University, Nanjing 210093, China}
\author{Zi Yang Meng}
%\email{zymeng@iphy.ac.cn}
\affiliation{Beijing National Laboratory for Condensed Matter Physics, Institute of Physics, Chinese Academy of Sciences, Beijing 100190, China}
\affiliation{Department of Physics and HKU-UCAS Joint Institute of Theoretical and Computational Physics, The University of Hong Kong, Pokfulam Road, Hong Kong, China}
%\affiliation{Songshan Lake Materials Laboratory , Dongguan, Guangdong 523808, China}
\author{Wei Li}
\affiliation{School of Physics, Beihang University, Beijing 100191, China}
\affiliation{International Research Institute of Multidisciplinary Science, Beihang University, Being 100191, China}
\author{Youguo Shi}
\email{ygshi@iphy.ac.cn}
\affiliation{Beijing National Laboratory for Condensed Matter Physics, Institute of Physics, Chinese Academy of Sciences, Beijing 100190, China}
\affiliation{Center of Materials Science and Optoelectronics Engineering, University of Chinese Academy of Sciences, Beijing 100049, China}
\affiliation{Songshan Lake Materials Laboratory, Dongguan, Guangdong 523808, China}
\author{Shiliang Li}
\email{slli@iphy.ac.cn}
\affiliation{Beijing National Laboratory for Condensed Matter Physics, Institute of Physics, Chinese Academy of Sciences, Beijing 100190, China}
\affiliation{School of Physical Sciences, University of Chinese Academy of Sciences, Beijing 100190, China}
\affiliation{Songshan Lake Materials Laboratory, Dongguan, Guangdong 523808, China}
\begin{abstract}
Strong spin-orbital-coupling magnetic systems with the honeycomb structure can provide bond-directional interactions which may result in Kitaev quantum spin liquids and exotic anyonic excitations. However, one of the key ingredients in real materials\---disorders\---has been much less studied in Kitaev systems. Here we synthesized a trigonal SrIr$_2$O$_{6-\delta}$ with $\delta \approx 0.25$, which consists of two-dimensional honeycomb Ir planes with edge-sharing IrO$_6$ octahedra. First-principles computation and experimental measurements suggest that the electronic system is gapped, and there should be no magnetic moment as the Ir$^{5+}$ ion has no unpaired electrons. However, significant magnetism has been observed in the material, and it can be attributed to disorders that are most likely from oxygen vacancies. No magnetic order is found down to 0.05 K, and the low-temperature magnetic properties exhibit power-law behaviors in magnetic susceptibility and zero-field specific heat, and a single-parameter scaling of the specific heat under magnetic fields. These results provide strong evidence for the existence of the random singlet phase in SrIr$_2$O$_{6-\delta}$, which offers a different member to the family of spin-orbital entangled iridates and Kitaev materials.
\end{abstract}

\maketitle

A two-dimensional (2D) honeycomb magnetic system may host a so-called Kitaev quantum spin liquid (QSL) if the magnetic interactions are strongly bond-dependent \cite{Kitaev2006,Baskaran2007,Feng2007}. Many iridates and $\alpha$-RuCl$_3$ have raised great interest recently, as the edge-sharing IrO$_6$ or RuCl$_6$ octahedra provides the unique building block of such bond-directional Kitaev interactions \cite{Jackeli2009,Chaloupka2010,RauJG14,Yamaji2014,Ye2012,Liu2011,Choi2012,Williams2016,Takayama2015,Kitagawa2018,Bahrami2019,BanerjeeA17,TakagiH19}. Although many of them show long-range magnetic ordering due to the presence of other types of interactions, such as Heisenberg and off-diagonal exchanges \cite{Chaloupka2010,RauJG14,Yamaji2014}, possible QSL states may be still realized through tuning magnetic field, pressure and chemical modification \cite{Kitagawa2018,Bahrami2019,BaekSH17,WangZ17,KasaharaY18,HermannV18}. However, despite great efforts and progress made, theoretical discussions and experimental verifications of the effects of disorder and randomness in the electronic and magnetic states of Kitaev materials are much less studied \cite{YadavR18,LiY18,KnolleJ19,ChoiYS19,YamadaMG20,BahramiF21,WHKao2102}, even though they are inevitable in real materials. 

Strong randomness can results in an interesting phase dubbed the random singlet phase (RSP), where the spins across arbitrary distances are able to form singlets due 
to the random distribution of antiferromagnetic (AFM) interactions \cite{LeeP85}.  
A crucial feature of a RSP is that its heat capacity $C$ and magnetic susceptibility $\chi$ must obey 
single-parameter scaling behaviors at low temperatures, such as both $C(T,H=0)/T$ and $\chi(T)|_{H\rightarrow 0}$ 
show power-law temperature dependence with a negative exponent between -1 and 0, and $C(H,T) \propto H^{-\gamma}F(T/H)$. The RSP has been well illustrated 
theoretically in one dimension (1D) \cite{DasguptaC80,TakemoriT82,FisherDS94}. The theoretical framework 
was extended to 2D and three-dimensional random spin-1/2 quantum magnets~\cite{BhattLee1982,Paalanen1988,Kimchi2018PRX,Kimchi2018NatComm}, but the problem cannot be solved 
in an analytically controlled manner. Only in recent years has unbiased quantum Monte Carlo seemed to reveal the 
existence of a 2D RSP for the square-lattice spin-1/2 quantum spin system with random interactions \cite{LiuLu2018}. It should be noted that these studies have not involved spin-orbital couplings (SOCs), which are fundamentally present in the above Kitaev materials.

Experimentally, the RSP has long been found in doped semiconductors, such as Si:P and CdS \cite{KummerRB78,Andres1981}. The essence for the existence of the RSP in these systems is that the spins are randomly distributed in space in essentially a continuum way, with the exchange coupling $J$ between any two sites falling off exponentially with distance $r$. This "continuum" is because the range of $J$ is set by the radius $a$ of the donor electron, which is much greater than the host lattice spacing. Recently, many frustrated magnetic systems showing no magnetic order have attracted a lot of interest, including iridates, as they may host the QSLs~\cite{Kitagawa2018,MaZhen2018,Klanjsek2017,Watanabe2018,MustonenO18a,Kitagawa2018,Bahrami2019}. Alternately, the picture of RSP also emerges as a very good guess to explain experimental observations for some of these magnets~\cite{Kimchi2018PRX,Kimchi2018NatComm,LiuLu2018,HongW21}, especially considering that these materials are very different in composition and microscopic magnetic interactions but surprisingly exhibit part of the single-parameter scaling behaviors described above. 

In this work we show that the RSP may also exist in a newly discovered strong SOC iridate, SrIr$_2$O$_{6-\delta}$, where Ir$^{5+}$ ions form a 2D honeycomb lattice. While Ir$^{5+}$ itself should have zero spin number, the oxygen vacancies in the material give rise to magnetic moments. Both the magnetic susceptibility and specific heat show scaling behaviors expected for the RSP, but the exponents are slightly different for these two properties, which indicates strong SOC effects. Our experimental observations are further supplemented by state-of-art first-principles calculations.

\begin{table}
  \centering
    \begin{tabular}{ccccccc}
        \hline
             & Site & $x$ & $y$ & $z$ & Occupancy & $U$ (\AA$^2$)\\
        \hline
        Sr & 1$a$ & 0.0000 & 0.0000 & 0.0000 & 1 & 0.0079(6) \\
        Ir & 2$d$ & 0.3333 & 0.6667 & 0.5000 & 1 & 0.0127(10) \\
        O & 6$k$ & 0.378(3) & 0.378(3) & 0.296(4) & 1 & 0.015(3) \\
        \hline
    \end{tabular}
  \caption{Nuclear structure parameters of SrIr$_2$O$_6$ with the space group of $P\bar{3}1m$ (No. 162): $a = b = 5.263(3)$ \AA, $c = 5.222(3)$ \AA, $\alpha = \beta = 90 ^{\circ}$, $\gamma = 120 ^{\circ}$. $R$: 0.048. $S$: 1.215.}
  \label{tab:table1}
\end{table}

Single crystals of SrIr$_2$O$_{6-\delta}$ were synthesized by mixing Sr(NO$_3$)$_2$ powders and Na$_2$IrO$_3$ single crystals, which were kept at 400$^\circ C$ for 10 h. The product was washed by deionized water many times and pure samples were collected after drying at 80$^\circ C$. The final crystals typically have hexagonal shape, with an in-plane size of about 1 mm and thickness of about 0.2 mm. The resistivity and heat capacity were measured by a Physical Property Measurement System (Quantum Design) with the dilution refrigerator option, and the magnetic susceptibility was measured by a Magnetic Property Measurement System (MPMS, Quantum Design) and a MPMS with the He-3 option. The crystal structure was determined by the single-crystal x-ray diffraction (SC-XRD). Scanning transmission electron microscopy (STEM) observations were performed in the JEOL ARM200F equipped with double aberration correctors and operated at 200 kV.  X-ray photoelectron spectroscopy (XPS) was measured on an AXIS Ultra system (Shimadzu, Japan) with an Al $K_{\alpha}$ radiation ($h\nu$ = 1486.6 eV). All binding energies were calibrated against the C 1s peak (284.8 eV), which arises from surface adventitious carbon. 
The first-principles calculations were carried out based on density-functional theory (DFT) using the projector augmented wave method as implemented in Vienna \textit{ab initio} simulation package~\cite{vasp,PAW}.

\begin{figure}[tbp]
\includegraphics[width=\columnwidth]{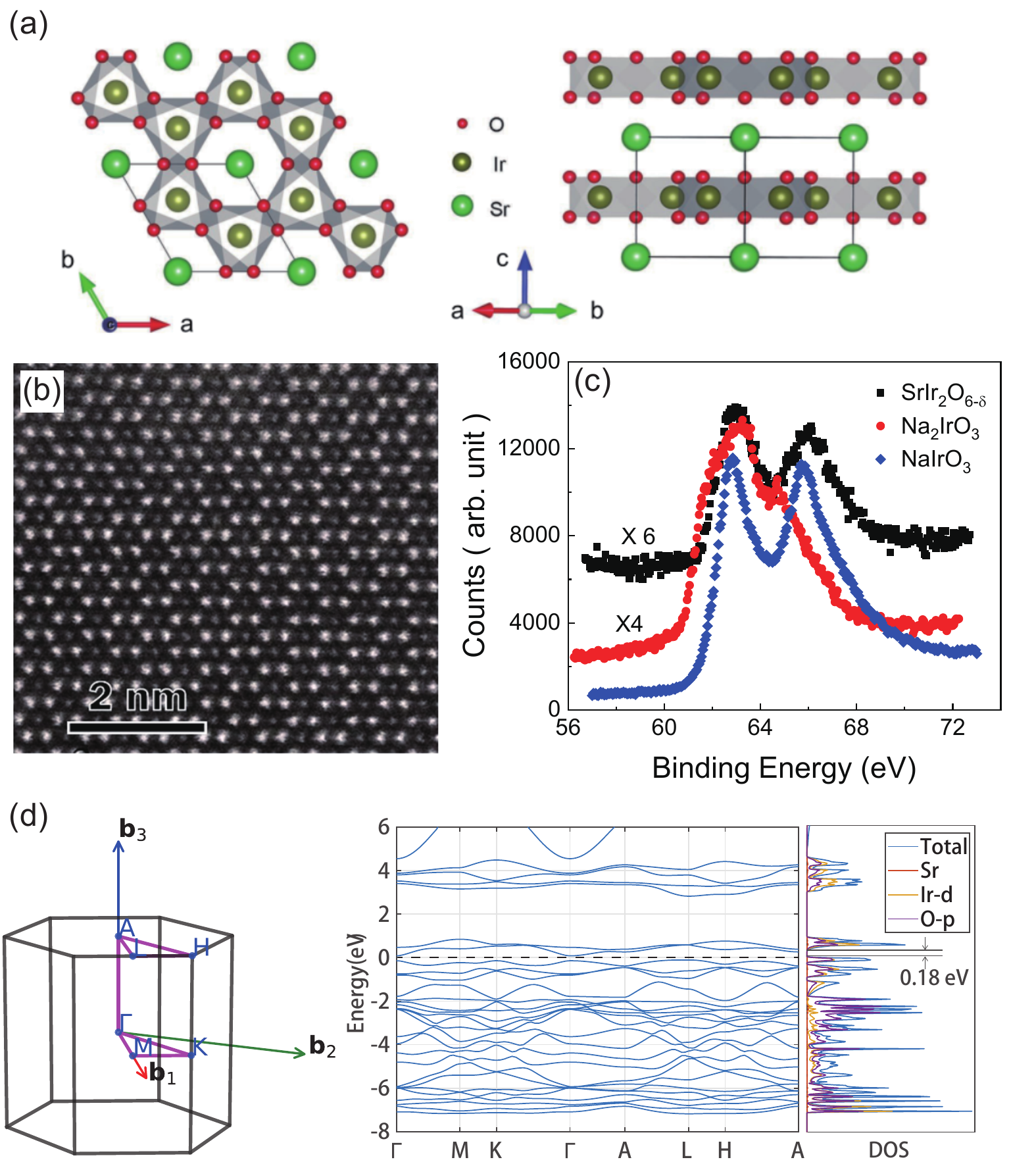}
 \caption{(a) The crystal structure of SrIr$_2$O$_{6-\delta}$ for the top view (left) and the side view (right). The edge-shared IrO$_6$ octahedra are shown. (b) The STEM image taken along the $c$-axis direction. According to the SC-XRD results, the light and dark dots correspond to the position of Ir and Sr, respectively.  (c) XPS spectra for SrIr$_2$O$_{6-\delta}$, Na$_2$IrO$_3$, and NaIrO$_3$. Both peaks come from the iridium ions. (d) (left) The first Brillouin zone for SrIr$_2$O$_6$. (right) Band structure and projected DOS of SrIr$_2$O$_6$. The horizontal dashed line represents the Fermi energy $E_F$.}
 \label{fig:figure1}
\end{figure}

Figure~\ref{fig:figure1} (a) gives the crystal structure of SrIr$_2$O$_{6-\delta}$ (space group $P\bar{3}1m$), where Ir$^{5+}$ ions with edge-shared IrO$_6$ octahedra form the honeycomb layers separated by the Sr$^{2+}$ ions. SrIr$_2$O$_{6-\delta}$ can be viewed as alternately stacking structure of layered IrO$_3$ and Sr atom layer. The detailed information for the crystal structure determined by the SC-XRD is given in Table~\ref{tab:table1}. Figure~\ref{fig:figure1} (b) shows a STEM image taken along the $c$-axis direction, which shows a perfect honeycomb arrangement of iridium ions. To obtain the valence of iridium in SrIr$_2$O$_{6-\delta}$, we compare its XPS spectra with those of NaIrO$_3$ and Na$_2$IrO$_3$ as shown in Fig.~\ref{fig:figure1} (c), which are known to have 5+ and 4+ iridium ions, respectively. It is clear that the spectra of SrIr$_2$O$_{6-\delta}$ is close to that of NaIrO$_3$, which suggests that the valence of iridium ions is mostly 5+. Comparing the integrated intensities of the two peaks shows that they are almost the same for NaIrO$_3$, but for SrIr$_2$O$_{6-\delta}$, the one at higher energy is about 11\% smaller than that at lower energy. Supposing that this difference is due to the existence of Ir$^{4+}$ whose spectra follow the one for Na$_2$IrO$_3$, we estimate that the content of Ir$^{4+}$ is about 25\%. Since no obvious disorders have been found in the SC-XRD data, we suspect that the origin of Ir$^{4+}$ in our system is associated with the presence of oxygen vacancies, which are rather insensitive to the x-ray diffraction. The amount of Ir$^{4+}$ corresponds to about 4\% oxygen vacancies ($\delta \approx$ 0.04) supposing one oxygen vacancy creates two Ir$^{4+}$. 

We calculated the band structure of SrIr$_2$O$_6$ by the DFT+U method~\cite{Anisimov1991,LDAU,dudarev1998} to understand its basic electronic properties. An effective exchange parameter $U_{eff}$ = 1.5 eV was found to quantitatively explain the experimental results as shown below and is also consistent with previous reports on some other iridates \cite{KimBJ08,DuL13}. The first Brillouin zone , the band structure and the density of states (DOS) are shown in Fig.~\ref{fig:figure1}(d). Our calculations show that the states from -8 to 6 eV are dominated by Ir-$5d$ and O-$2p$ orbitals and the splitting gap between $t_{2g}$ and $e_g$ is about 2 eV. Moreover, the system is an insulator with a band gap $\sim$ 0.18 eV. As already known \cite{Jackeli2009,Chaloupka2010,RauJG14,Yamaji2014}, the octahedral crystal field of IrO$_6$ and strong SOC effects result in a high-energy $j$ = 1/2 doublet and a low-energy $j$ = 3/2 quartet. Therefore, for Ir$^{5+}$, the system has the fully occupied $j=3/2$ and empty $j = 1/2$ states and thus should not exhibit magnetism, as shown in NaIrO$_3$ and Sr$_3$CaIr$_2$O$_9$ \cite{WallaceDC15}.

\begin{figure}[tbp]
\includegraphics[width=\columnwidth]{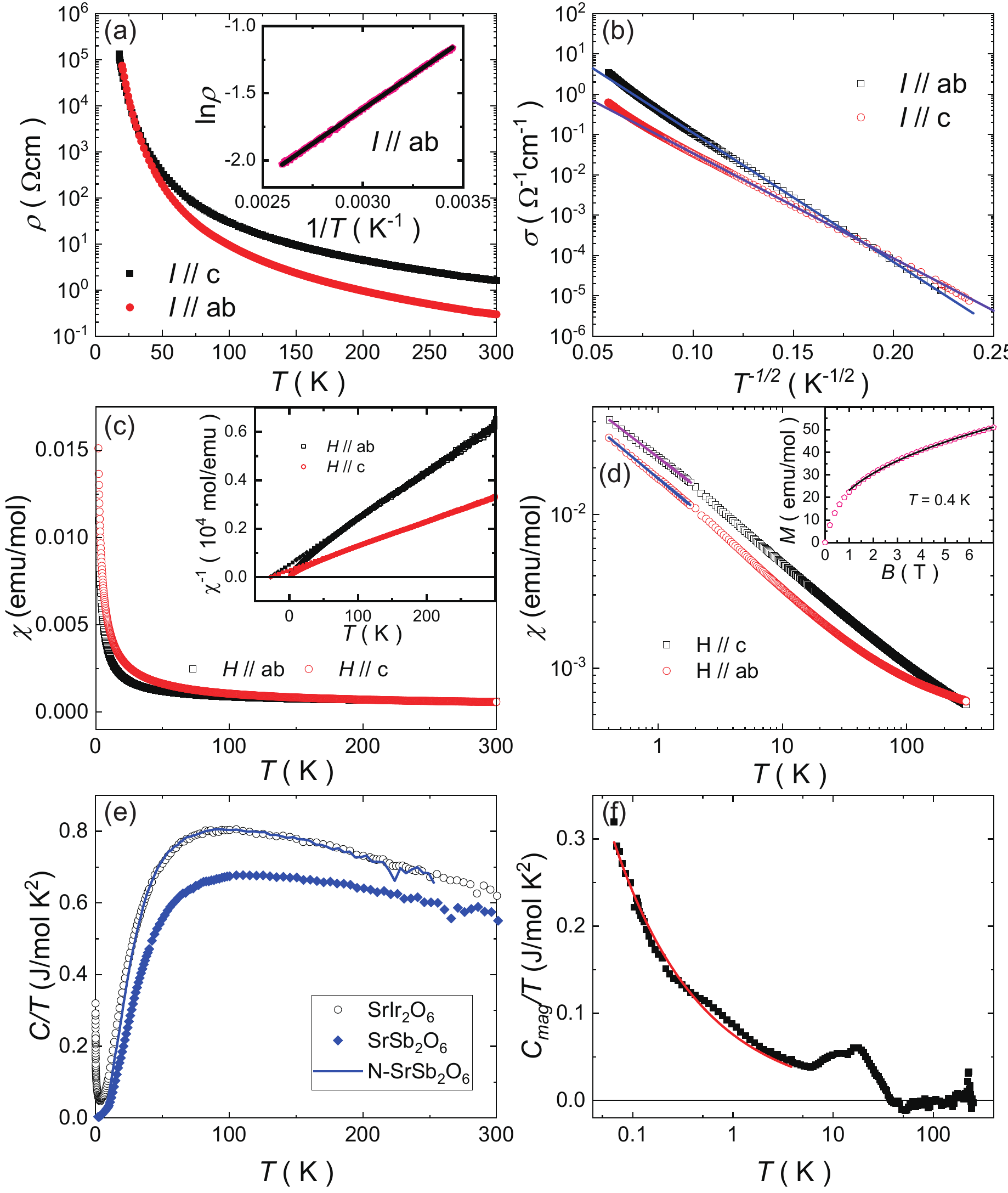}
 \caption{(a). The temperature dependence of resistivity $\rho$. The inset shows that the high-temperature data can be fitted by a gap function as described in the main text. The solid line is fitted by the gap function. (b) The $T^{-1/2}$ dependence of conductivity $\sigma$. The dashed lines are fitted results for the low-temperature data as described in the text. (c) The temperature dependence of the magnetic susceptibility $\chi$ = $M/H$ at 0.1 T. The inset shows the temperature dependence of $\chi^{-1}$ after subtracting a constant background. The dotted lines are the fitted results of the linear function. (d) The temperature dependence of $\chi$ with the log-log scale. The solid lines are fitted by the $T^{-\alpha}$ function. The data below 1.8 K have been normalized to those above. The inset shows the $M-H$ curve at 0.4 K for $H // c$ with the data above 1 T fitted by the $B^{1-\alpha}$ function (solid line) \cite{Kimchi2018NatComm}. (e) The temperature dependence of the specific heat. The solid line shows normalized specific heat of SrSb$_2$O$_6$ as described in the text where $T$ has been replaced by $bT$. (f) The temperature dependence of the magnetic specific heat. The solid line is fitted by $AT^{-0.5}$, where $A$ is the temperature-independent parameter.}
 \label{fig:figure2} 
\end{figure}

The temperature dependence of the resistivity is consistent with the above calculated results, as shown in Fig.~\ref{fig:figure2} (a). The high-temperature data ( $>$ 300 K ) shows a gap behavior, i.e. $\rho \propto \exp(-E_g/2k_BT)$, as shown in the inset of Fig.~\ref{fig:figure2} (a). The fitting gives a gap value of $\sim$ 0.175 eV, which is close to the calculated one. At low temperatures, the temperature dependence of the resistivity can be described by the variable-range-hopping (VRH) theory \cite{ShklovskiiBI84}, where the low-temperature conduction is determined by the activated hopping between the localized states near the Fermi level. Accordingly, the conductance $\sigma$ is proportional to $\exp[-(T_0/T)^{1/2}]$, where $T_0$ is a characteristic temperature, as shown by the dashed lines in Fig.~\ref{fig:figure2} (b). The power of $1/2$ may come from the existence of the Coulomb gap in the localized states due to electron-electron interactions \cite{ShklovskiiBI84}. Fitting the data with this expression gives the values of $T_0$ for the current within the $ab$ plane and along the $c$ axis as about 5400 and 3600 K, respectively. According to the VRH theory \cite{ShklovskiiBI84}, $T_0 = 2.8e^2/4\pi\epsilon\epsilon_0ak_B$, where $e$, $\epsilon_0$, $\epsilon$, and $k_B$ are the electron charge, vacuum permittivity, relative permittivity, and Boltzmann constant, respectively. Here $a$ is the radius of localized states near the Fermi level, so we have $\epsilon a$ = 8.7 nm within the $ab$ plane and 12.8 nm along the $c$ axis.  These large values of radius closely resemble those of donor electrons in doped semiconductors \cite{KummerRB78,Andres1981} and are consistent with the random but long-ranged interactions between the magnetic moments.

Figure~\ref{fig:figure2} (c) gives the temperature dependence of the magnetic susceptibility $\chi$. The data above 50 K can be fitted by the Curie-Weiss function with $\chi = C/(T-\theta)+\chi_0$, where $A$, $\theta$, and $\chi_0$ are the Curie constant, Curie temperature, and backgrounds, respectively. Along both directions, $\theta$ is about -27 K, suggesting the interactions are dominated by the AFM exchanges. The effective moments $\mu_{eff}$ are 0.46 and 0.63 $\mu_B$ for the field within the ab plane and along the $c$-axis, respectively. These small values are consistent with the estimated amount of Ir$^{4+}$ with $S$ = 1/2, as shown above. The value of $\chi_0$ for the field within the $ab$ plane and along the $c$ axis is 4.5 $\times 10^{-4}$ and 2.8 $\times 10^{-4}$ emu/mol, respectively, suggesting a large Van Vleck contribution. At low temperatures, $\chi$ exhibits a power-law temperature dependence, i.e., $\chi \propto T^{-\alpha}$, as shown in Fig.~\ref{fig:figure2} (d). Fitting the data below 2 K gives $\alpha$ equal to 0.62 and 0.67 for $H // c$ and $H // ab$, respectively. Alternately, fitting on the high-field $M-H$ curve shows that $\alpha$ = 0.6, consistent with the result from $\chi(T)$.

Figure~\ref{fig:figure2} (e) shows the specific heats of SrIr$_2$O$_{6-\delta}$ and its isostructural nonmagnetic compound SrSb$_2$O$_6$ at zero field. The phonon part of the specific heat in the former, $C^{Ir}_{ph}$, can be obtained from that of the latter, $C^{Sb}$, by simply introducing a renormalization factor $b$ for the temperature, i.e., $C^{Ir}_{ph}(T)$ = $C^{Sb}(bT)$. Here $b$ is calculated to be about 0.841 based on the molar mass of the two samples with {\it no} manual adjustment \cite{BouvierM91}, i.e.,
\begin{equation}
b = \left[\frac{M_{Sr}^{3/2}+2M_{Ir}^{3/2}+6M_{O}^{3/2}}{M_{Sr}^{3/2}+2M_{Sb}^{3/2}+6M_{O}^{3/2}}\right]^{1/3},
\label{bofC}
\end{equation}
\noindent where $M_{Sr}$, $M_{Ir}$, $M_{Sb}$, and $M_{O}$ are the atomic mass of Sr, Ir, Sb, and O, respectively.
The overlap at temperatures larger than about 50 K suggests that this method correctly obtain the phonon contribution to the specific heat in SrIr$_2$O$_{6-\delta}$. The magnetic specific heat can thus be obtained by subtracting this part from the total specific heat, as shown in Fig.~\ref{fig:figure2} (f). The low-temperature specific heat roughly show $T^{-0.5}$ behavior, but there are small features that deviate from this simple temperature dependence, which is reasonable for a RSP since the power-law DOS of the randomness is an approximation \cite{Kimchi2018PRX,Kimchi2018NatComm}.  A broad hump also appears around 15 K, which typically marks the temperature where the short-range correlations of spins for a 2D system are built up and is consistent with the absolute value of the Curie temperature $\theta$ obtained above. Accordingly, the magnetic entropy is about 1.6 J/mol K, which is about 0.27Rln2. This value is very close to the estimated content of Ir$^{4+}$. Moreover, the effective moment from these spins will be 0.47 $\mu_B$ (for $g$ = 2), which is also close to the experimental values obtained from the magnetic-susceptibility measurements. Of course, the actual case for the moments may be more complicated as discussed later in our theoretical calculations.

\begin{figure}[tbp]
\includegraphics[width=\columnwidth]{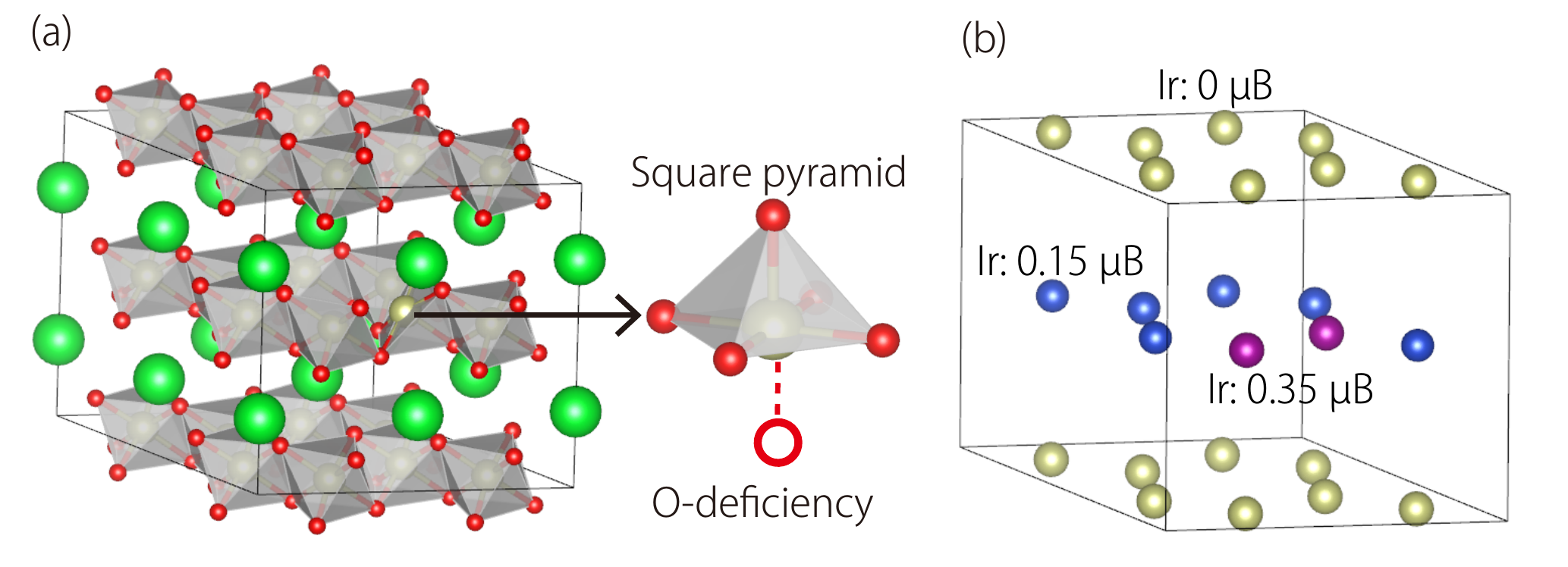}
\caption{(a) The crystal structure of SrIr$_2$O$_6$ with one oxygen vacancy in a 2$\times$2$\times$2 supercell. (b) Schematic view of Ir framework with magnetic moments (in $\mu_B$) obtained from first-principles calculation.}
\label{fig:figure3}
\end{figure}

As suggested by the VRH behavior of the resistivity, there should have disorders in our sample that result in the localized states near the Fermi level. These disorders are most likely the oxygen vacancies, whose content is about 4\% as discussed above. As shown in Fig.~\ref{fig:figure3}, with oxygen vacancies, the crystal field surrounding each Ir would be transformed from octahedron to square pyramid, so that the $t_{2g}$ orbitals may further split into a lower doublet (including $d_{xz}$ and $d_{yz}$) and a higher singlet $d_{xy}$. Two local Ir$^{5+}$ ions becomes Ir$^{4+}$ with $5d^5$ configuration, such that the $d_{xz}$ and $d_{yz}$ orbitals are fully filled and the $d_{xy}$ orbital may be half-filled, introducing local magnetic moments.

We performed the first-principle calculation to reveal such scenario. In Fig.~\ref{fig:figure3} (a), we consider a $2\times 2 \times 2$ supercell with  one oxygen vacancy, such that two Ir atoms connected by the oxygen atom become  square pyramid surrounding. Indeed, magnetic moment emerges mainly for Ir$^{4+}$ ions and their in-plane neighbors [Fig.~\ref{fig:figure3}(b)].  In fact, we may even quantitatively compare the experimental and theoretical results. The effective content of the oxygen vacancies is 4.16\% since an oxygen vacancy has no effect on adjacent layers. The induced total moment is 1.88 $\mu_B$, corresponding to 0.94 $\mu_B$/Ir$^{4+}$. This value is very close to the magnetic moment of a spin 1/2 with $g$ = 2 and thus gives rise to an effective moment of about 0.43 $\mu_B$/Ir for with 25\% Ir$^{4+}$, very close to the experimental values. The small values of the effective moments and magnetic entropy have also been observed in many iridium oxides with 5$d^4$ configurations, which have raised many interests and debates~\cite{KhaliullinG13,CaoG14,MeeteiO15,DeyT16,AbhishekN16,CorredorLT17,FuchsS18,KuschM18,AbhishekN18,ParamekantiA18}. Here in SrIr$_2$O$_{6-\delta}$, our analysis based on oxygen vacancies gives a natural explanation of the origin of its magnetism.

\begin{figure}[tbp]
\includegraphics[width=\columnwidth]{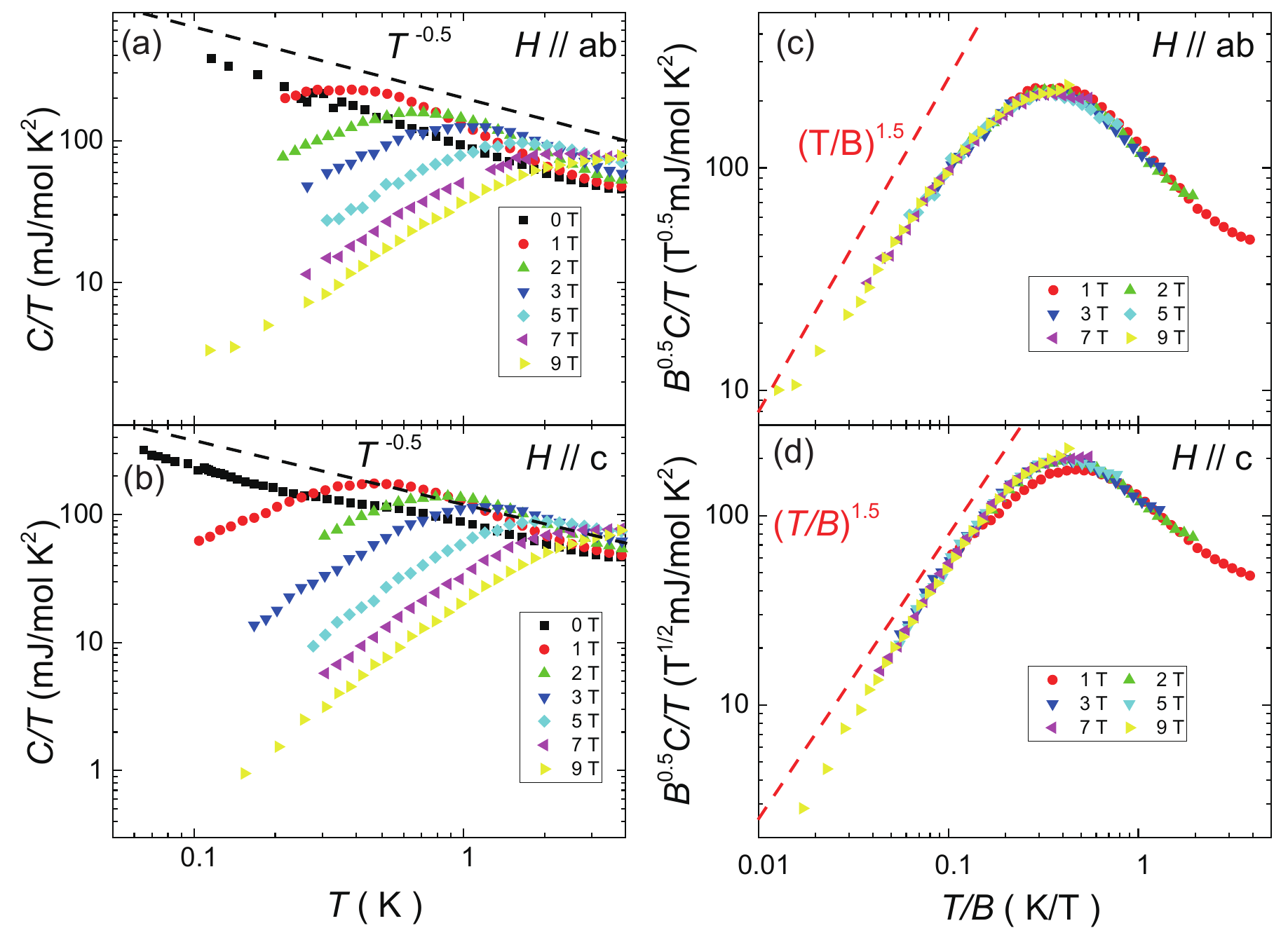}
\caption{(a), (b) The temperature dependence of $C/T$ for $H // ab$ and $H // c$, respectively. The dashed lines are guides to the eye. Here the phonon contributions are not considered as they are negligible in this temperature range. (c), (d) The scaling behavior of $C(T,H)$ for $H // ab$ and $H // c$, respectively. The dashed lines are guides to the eye.}
\label{fig:figure4}
\end{figure}

The ground states of the magnetic system in SrIr$_2$O$_{6-\delta}$ can be revealed by the low-temperature specific heats as shown in Fig.~\ref{fig:figure4} (a) and (b). At zero field, the $C/T$ shows $T^{1-\gamma}$ behavior at low temperatures with $\gamma$ equal to about 0.5. The suppression of the specific heat at low temperatures by the magnetic field for $H // c$ is more significant than that for $H // ab$, which is consistent with $\chi_c > \chi_{ab}$. Figures~\ref{fig:figure4} (c) and (d) show that all data fall onto one curve when they are plotted as $B^{0.5}C/T$ vs $T/B$. In other words, a single-parameter scaling can describe the field effects on the specific heats in SrIr$_2$O$_{6-\delta}$ \cite{Kimchi2018PRX,Kimchi2018NatComm}. 

Based on the above results, we conclude that the magnetic ground state in SrIr$_2$O$_{6-\delta}$ is the RSP, which is the first kind in Kitaev materials.  
The major evidences for the existence of the RSP include that the magnetic susceptibility and specific heat at zero field follows power-law temperature dependence, i.e., $\chi_{H\rightarrow0} \propto T^{-\alpha}$ and $C(H=0) \propto T^{1-\gamma}$, and more importantly, the scaling behavior of the specific heat under field, i.e., $C/T \propto H^{-\gamma} F_q(T/H)$, which have been treated as the fundamental properties of the RSP \cite{Kimchi2018PRX,Kimchi2018NatComm}. We note the comparison between the doped semiconductors showing the RSP \cite{KummerRB78,Andres1981} and the SrIr$_2$O$_{6-\delta}$ system also bears interesting similarities. Both the electronic systems are gapped while significant local states exist near the Fermi levels. In the former, the local states come from random donor electrons, whose large radius $a$ results in the long-range exchange coupling $J$ that leads to the RSP. In SrIr$_2$O$_{6-\delta}$, the radii of local electrons near the Fermi level, which should be associated with oxygen vacancies, are also large as suggested by the VRH behavior. These similarities may explain the origin of the RSP in SrIr$_2$O$_{6-\delta}$. Another factor for the formation of singlets may be the tendency of dimerization in this compound, which competes with magnetic order and Kitaev spin liquid states as in LiRuO$_3$ and $\alpha$-Li$_2$IrO$_3$ under pressure \cite{ParkJ16,HermannV18}. Nevertheless, the origin of RSP in SrIr$_2$O$_6$ can only be answered in the future by establishing a random spin model, and find the quantum many-body states therein by solving this model.

A unique feature for the RSP in SrIr$_2$O$_6$ is that the values of $\alpha$ in the magnetic-susceptibility scaling and $\gamma$ in the specific-heat scaling are not the same, which contradicts with the simple predicted properties of a RSP \cite{Kimchi2018PRX,Kimchi2018NatComm}. We note that there is strong SOC in our system, which may give rise to this discrepancy. The effects of SOC can be indeed seen in the low-temperature behavior of $B^{\gamma}C/T$, which shows $(T/B)^{q}$ dependence [Figs. \ref{fig:figure4}(c) and \ref{fig:figure4}(d)]. Theoretically, it has been argued that the value of $q$ is associated with the SOC and antisymmetric Dzyaloshinskii-Moriya spin exchanges \cite{Kimchi2018NatComm}. Here the values of $q$ are about 1.5 for both field directions, which may come from combinations of different cases. Moreover, the fact that there should be bond-directional interactions in this material may also have significant effects in its magnetic properties \cite{KhaliullinG13}.  
With all these findings, our results offer another honeycomb iridate member to the family of iridates and Kitaev materials, and reveal the fact that the SOC quantum magnets, can not only give rise to exotic magnetic order, quantum spin liquid but also the RSP. In future, it will be of great interests to the investigate the possible competition and collaboration of these phases in the iridates and Kitaev materials.

\acknowledgments

This work is supported by Beijing Natural Science Foundation (Grants No. Z180008), the National Key Research and Development Program of China (Grants No. 2017YFA0302900, No. 2016YFA0300500, No. 2016YFA0300604,2020YFA0406003), the National Natural Science Foundation of China (Grants No. 11874401, No. 11961160699, No. U2032204, No.  11774399, No. 11974036), the K. C. Wong Education Foundation (Grants No. GJTD-2020-01, No. GJTD-2018-01), the Strategic Priority Research Program (B) of the Chinese Academy of Sciences (Grants No. XDB33000000),  and the RGC of Hong Kong SAR of China (Grant Nos. 17303019 and 17301420).

P. S., K. Z. and F. Y. contributed equally to this work.

\end{document}